**CONSTRUCTION OF A PZT SENSOR NETWORK FOR LOW AND HYPERVELOCITY IMPACT DETECTION.** J. A. Carmona, M. Cook, M. Cooper, J. Schmoke, J. Reay, L. Matthews and T. Hyde, CASPER (Center for Astrophysics, Space Physics and Engineering Research), Baylor University, Waco, TX 76798 E-mail: Truell_Hyde@baylor.edu.

**Introduction:** Orbital debris is a constraint on the long-term health of any spacecraft and must be considered during mission planning. Varying mechanisms have been proposed to quantify the problem. Assessment of orbital debris employing ground-based methods such as radar can help determine where debris clouds are located as well as their density or orbital trajectory. Such data is invaluable to computer simulations and can allow predictions of the debris environment over specific time periods [1]. Accurate in-situ data is essential as well with various types of sensors designed to detect orbital debris impacts employed on space missions since the 1950's [2]. One of the most common of these is the PZT (piezoelectric lead zirconate titanate) which is often used in-situ to measure the momentum of a particle at the time of impact. This paper will discuss a multiple PZT sensor system capable of determining both impactor momentum and location currently in development within CASPER.

**Experiment:** In order for any PZT sensor system to measure both the momentum and location of an impact, the plate to which the PZT is attached must first be carefully mapped. This mapping process must be completed for each PZT individually as well as for the final sensor network and is accomplished in two steps. First a low-velocity mapping of the plate is accomplished using a drop tower. Once this is completed, a second higher-velocity mapping employing a light gas gun (LGG) is conducted. During the first part of the mapping process, the particle is dropped from differing heights, where gravity is the only force acting on it. As a result, the drop height can be used to calculate the velocity of the particle when it strikes the plate using equation (1)

$$V = (2gh)^{1/2} \qquad (1)$$

where h is the drop height and g is the acceleration of gravity. In the LGG, velocities are measured directly employing laser fan diagnostic devices (developed specifically for the LGG at CASPER) that measure the particle travel time between two known positions along the beamline. In both phases, once known velocities have been established for all regions of the plate, it can then be mapped by calculating the relationship between the momentum delivered at the time of impact and the electrical response as measured at the sensor. This ratio is called the sensitivity, S, and has units of V/N·s. The sensitivity of the plate varies over its face and is dependent (among other things) on composition, the manner in which it is clamped, and the location(s) of the PZT(s). As mentioned, a proper mapping of the plate requires knowing the change in sensitivity across all regions of the plate for each PZT sensor as well as the final sensor network. Once this is known, data collected from the plate should enable the momentum and location of a specific particle impact to be identified.

*Witness plate.* During data collection, a six-inch stainless steel circular flange is attached to one port of a six-way vacuum chamber (which is connected to the LGG) and employed as the impact plate. This arrangement has the advantage that the PZT's attached to the back of the plate are now outside the vacuum environment providing optimal accessibility and vastly improved experiment set up times. PZT's are attached to the impact plate using a custom clamp structure that screws into the plate. The clamp structure employs three different mounting locations to hold the PZT to the back of the plate while assuring uniform pressure across the PZT surface.

*PZT network.* Once the impact plate with a single PZT attached is mapped using both the drop tower and the LGG, a second PZT is added. Clamping of the new PZT is accomplished in the manner described above and a PZT identical to the first is used. Once a new PZT is added to the system, the plate is mapped again and then placed in the LGG for data collection. With one PZT attached to the back of the plate the expected sensitivity response (to first order) is that of a Gaussian distribution. For two PZT's, the expected response is two Gaussians in the same plane with three Gaussian distributions for three PZT's and so forth. The maximum for each of the Gaussians should be located directly over the PZT location on the plate. Thus given a sensor network of this sort, the impact location of a particle on the plate can be retrieved. Based on the magnitude of the sensitivity measured at the time of impact, the momentum of the projectile can also be estimated using the mapping data, which in turn helps determine the velocity of the projectile if the mass is known. However as mentioned, the above is only true to first order since for accurate results the procedure given above must be modified by taking the acoustical behavior of the plate into consideration. This is due to the fact that both nodes and antinodes form on the plate during impact, modifying the PZT response and often yielding results different from those expected in first order.

**Results:** Figure 1 shows data taken employing the stainless steel plate with one PZT attached to it. The results show a distinct Gaussian profile with a maximum located at the center of the plate directly over the PZT location.

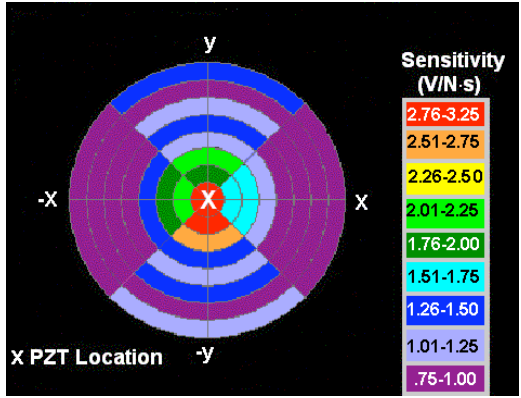

Fig. 1. Map of the stainless steel plate using one PZT.

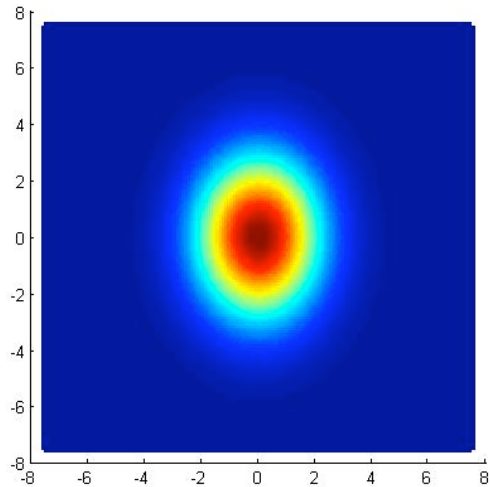

Fig. 2. Theoretical results for a mapping of the stainless steel plate (top view).

To simulate a multiple PZT network, a mathematical model was developed based on a normal distribution (as given by Eq. 2) with the variance determining the width of the Gaussian calculated using equation (3). This yields the elliptical two-dimensional Gaussian profile shown in Figures 2 and 3.

$$f(x,y) = \frac{175}{2*\pi*3.8*2.2} \exp-\left[\frac{x^2}{2*3.8} + \frac{y^2}{2*2.22}\right] \quad (2)$$

$$\sigma^2 = \frac{1}{n-1}\sum_{j=1}^{n}(x_j - \bar{x})^2 \quad (3)$$

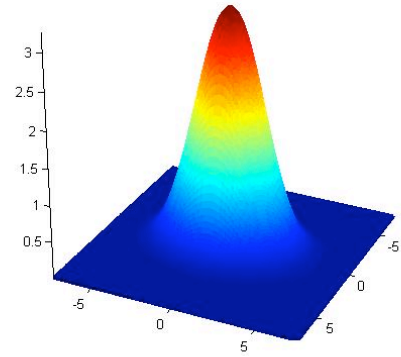

Fig. 3. Side view of a theoretical mapping of the stainless steel plate.

Preliminary experimental data collected on the plate employing the LGG (as shown in Figures 4 and 5) appears to agree with this model indicating a response indicative of a Bessel function in the x direction and a Gaussian function in the y direction. These results will be discussed in an upcoming publication.

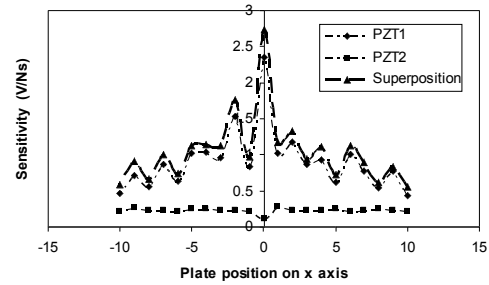

Fig. 4. Individual response for two PZT's overlaid with their combined response along the x-axis of the plate. (PZT's are located at the x = 0, -5 position.)

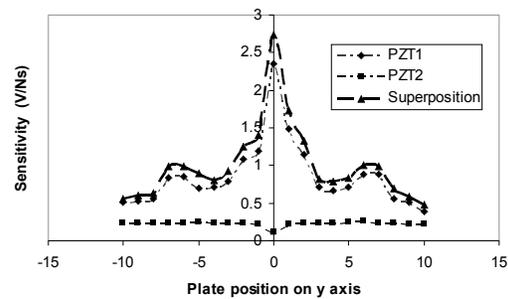

Fig. 4. Individual response of two PZT's overlaid with their combined response along the y-axis of the plate. (PZT's are located in the x = 0, -5 position).

**References:** [1] Corvonato E. et al (2001) *Internationla Journal of Impact Engineering,* 26, 115-128. [2] Alexander W. M. et al. (1965) *Science,* 149, 3689-3695.